\documentstyle[amssymb,aps,preprint]{revtex}

\tightenlines

\begin{document}
\title{Evidence for two-dimensional nucleation of superconductivity in MgB$_{2}$}
\author{A.S. Sidorenko$^{1,2}$, L.R. Tagirov$^{1,3},$ A.N. Rossolenko$^{4}$, V.V.
Ryazanov$^{4}$, R. Tidecks$^{1}$}
\address{$^{1}$Institut f\"{u}r Physik, Universit\"{a}t Augsburg, D-86159 Augsburg,
Germany\\
$^{2}$Institute of Applied Physics, 2028 Kishinev, Moldova\\
$^{3}$Kazan State University, 420008 Kazan, Russia\\
$^{4}$Institute of Solid State Physics, 142432 Chernogolovka, Russia}
\date{\today{}}
\maketitle
\draft

\begin{abstract}
According to the crystal structure of MgB$_{2}$ and band structure
calculations quasi-two-dimensional (2D) boron planes are responsible for the
superconductivity. We report on critical fields and resistance measurements
of 30 nm thick MgB$_{2}$ films grown on MgO single crystalline substrate. A
linear temperature dependence of the parallel and perpendicular upper
critical fields indicate a 3D-like penetration of magnetic field into the
sample. Resistivity measurements, in contrast, yield a temperature
dependence of fluctuation conductivity above T$_{c}$ which agrees with the
Aslamazov-Larkin theory of fluctuations in 2D superconductors. We consider
this finding as an experimental evidence of two-dimensional nucleation of
superconductivity in MgB$_{2}$.
\end{abstract}


\section{Introduction}

Recent discovery \cite{Nature} of a medium-temperature superconductivity in
magnesium diboride (MgB$_{2}$) raised questions about the origin and
properties of superconductivity in this compound. MgB$_{2}$ has a hexagonal
crystal structure with boron layers interleaved by magnesium layers. Due to
this layered structure, normal state electric transport, as well as
superconducting properties should be highly anisotropic. Band structure
calculations \cite{An,Kortus} indicate that electrons at the Fermi level are
predominantly derived from boron atoms. MgB$_{2}$ may be regarded as sheets
of metallic boron with strong covalent intralayer bonding, separated by Mg
layers with ionic interlayer B-Mg bonding. The strong B-B bonding induces
enhanced electron-phonon interaction, so that the superconductivity in MgB$%
_{2}$ is mainly due to the charge carriers in the boron planes.

Experimental investigations on single crystals and $c$-oriented epitaxial
and textured films (see, e.g., the review \cite{Buzea} and references
therein) give evidence for a highly anisotropic superconducting gap.
Measured critical magnetic fields usually show a pronounced anisotropy for $c
$-oriented films and single crystals \cite{Buzea}. Applying the anisotropic
Ginzburg-Landau model to these measurements, authors derive an effective
mass anisotropy for the charge carriers of $\gamma =m_{ab}/m_{c}\approx
0.15-0.3.$ Thus, band structure calculations and experimental measurements
strongly suggest that superconductivity nucleates at the
quasi-two-dimensional (quasi-$2D$) boron planes, and then extends through
the magnesium layers by a nanoscale proximity effect forming an anisotropic $%
3D$ superconducting state in the material.

In this Letter we present experimental evidence for $2D$ nucleation of
superconductivity in a  $3D$ magnesium diboride film. To demonstrate this,
we measured the temperature dependence of excess conductivity caused by
fluctuations above $T_{c}.$ If quasi-$2D$ boron planes are responsible for
the superconductivity, then the excess conductivity should exhibit $2D$
behavior although measured in a $3D$ sample. We found that the temperature
dependence of the excess conductivity agrees well with the Aslamazov-Larkin 
\cite{AL} theory of superconducting fluctuations in $2D$ superconductors.

\section{Results and discussion}

The MgB$_{2}$ films were prepared by DC magnetron sputtering on single
crystalline MgO (101) substrate according the procedure described in \cite
{Erm}. To compensate losses of magnesium due to its oxidation in plasma, a
composite target was used which contained MgB$_{2}$ and metallic magnesium
in approximately equal amounts. The Mg-MgB$_{2}$ pellet was sputtered in a
99,999\% purity argon atmosphere at a pressure of 3 Pa. The substrate
temperature during sputtering was held at 200 $^{0}$C and than raised to 600 
$^{0}$C for several seconds at the final stage. At this final {\it in situ}
annealing the plasma discharge was not switched off. X-ray studies revealed
a strongly textured (101)-oriented structure of our films. Films of 30 nm
thickness have been used for resistance and critical field measurements. 

The superconducting transition temperature, $T_{c},$ obtained by a
conventional four-terminal resistive method at zero field, was about 19.5K.
The upper critical fields parallel ($B_{c2\parallel }$) and perpendicular (B$%
_{c2\perp }$) to the film plane have been measured using the 7T
superconducting magnet of a ''MPMS XL Quantum Design'' SQUID magnetometer. 

The resistive transitions, $R(T),$ at constant perpendicular magnetic field
for one of the investigated samples are plotted in Fig.1. The transition
width at zero field, according to a (10-90)\% $R_{n}$ criterion, is about $%
0.3$K and slightly increases in stronger magnetic field to about 1K at 6T.
The temperature dependence of the critical fields is displayed in Fig.2. It
shows a pronounced anisotropy, $B_{c2\parallel }/B_{c2\perp }\simeq 1.5,$ in
accordance with previous studies, reporting values of $1.2-2$ for $c$%
-oriented films and 2.4-2.7 for single crystals \cite{Buzea}. The
temperature dependence of $B_{c2\parallel }(T)$ is linear except for
temperatures very close to the critical temperature. The $B_{c2\perp }(T)$
dependence starts nearly linear, and then increases more rapidly. We obtain
the Ginzburg-Landau coherence length, $\xi _{GL}(0),$ from the slopes of $%
B_{c2\parallel }(T)$ and $B_{c2\perp }(T)$ close to the critical temperature 
\cite{Tinkham}, 
\begin{equation}
\xi _{GL}(0)=\left[ -\left( dB_{c2}(T)/dT\right) \left( 2\pi T_{c}/\phi
_{0}\right) \right] ^{-1/2},  \label{eq0}
\end{equation}
where $\phi _{0}$ is the magnetic flux quantum. Calculations according to (%
\ref{eq0}) give $\xi _{GL\parallel }(0)=4.4$ nm, $\xi _{GL\perp }(0)=6.8$
nm, much less than the film thickness $d=30$ nm. The linear temperature
dependence of $B_{c2\parallel }(T)$ gives evidence that the film is three
dimensional with respect to superconductivity. Especially, the absence of a
square-root temperature dependence of $B_{c2\parallel }(T)$ clearly
demonstrates that the films show no indication for $2D$ superconductivity.
For the deviation of $B_{c2\perp }(T)$ from the linear behavior several
reasons may be responsible, such as: anisotropy of the energy gap, proximity
effect due to the weakly superconducting Mg interlayers in the MgB$_{2}$
compound \cite{Sid1,Sid2}. The nonlinear behavior of $B_{c2\perp }(T)$ is
commonly observed in MgB$_{2}$ (see Ref. \cite{Buzea}).

Using our data on the temperature dependence of the resistance at zero DC
magnetic field we calculate the fluctuation conductance by the relation \cite
{AL} 
\begin{equation}
\sigma ^{\prime }(T)=\frac{1}{R(T)}-\frac{1}{R_{n}}\propto
(T/T_{c}^{AL}-1)^{D/2-2},  \label{eq1}
\end{equation}
where $R_{n}$ is the normal state resistance, $D$ is the effective
dimensionality of a superconductor ($D=1,2,3$), and $T_{c}^{AL}$ is the
Aslamazov-Larkin critical temperature\cite{AL}. If the effective
dimensionality of superconducting fluctuations is $D=2$, then according to
Eq. (\ref{eq1}) we expect a linear dependence of the inverse excess
conductance on temperature: 
\begin{equation}
\left[ \sigma ^{\prime }(T)\right] ^{-1}=\frac{R_{n}}{\tau _{AL}}\frac{%
\left( T-T_{c}^{AL}\right) }{T_{c}^{AL}},  \label{eq2}
\end{equation}
with 
\begin{equation}
\tau _{AL}=C_{0}R_{n}^{\square },  \label{eq3}
\end{equation}
\begin{equation}
C_{0}=\frac{e^{2}}{16\hbar }=1.52\times 10^{-5}\text{ }\Omega ^{-1},
\label{eq4}
\end{equation}
where $R_{n}^{\square }$ is the normal state sheet resistance of the film. 

Fig. 3 shows the temperature dependence of the inverse excess conductance
normalized by the normal state conductance. The intersection of the linear
approximation by Eq. (\ref{eq2}) with the abscissa gives the critical
temperature $T_{c}^{AL}$, while the slope of the line provides the value of $%
\left( \tau _{AL}\right) ^{-1}$. In Fig. 3 we also show the fit according to
Eq. (\ref{eq1}) for the case $D=3.$ Obviously, the $2D$-fit describes the
experimental data much better then $3D$-fit. Using the experimental values
of the slope of the $2D$-fit we calculate the effective normal state sheet
resistance by Eqs. (\ref{eq3}) and (\ref{eq4}): 
\begin{equation}
R_{n%
\mathop{\rm eff}%
}^{\square }=\frac{\tau _{AL}}{C_{0}}=71\Omega /\Box .  \label{eq5}
\end{equation}
For a homogeneous $2D$ film this value should be identical to that one
obtained from the measurement of the resistance in the normal state above
the critical temperature, $R_{n\exp }^{\square }$. Since, however, $R_{n\exp
}^{\square }=1.105\Omega /\Box $ is about 60 times smaller than\ the
effective sheet resistance (\ref{eq5}) our sample must be a stack of a large
number of parallel boron sheets. The number of the sheets, $N_{%
\mathop{\rm eff}%
}$, which are involved in the sheet resistance of our MgB$_{2}$ film, may be
estimated as 
\begin{equation}
N_{%
\mathop{\rm eff}%
}=\frac{R_{n%
\mathop{\rm eff}%
}^{\square }}{R_{n\exp }^{\square }}=\frac{71}{1.105}\sim 64.  \label{eq6}
\end{equation}
In that case the effective interlayer spacing of the boron sheets can be
obtained from the film thickness, $d,$ as follows
\begin{equation}
d_{%
\mathop{\rm eff}%
}^{BB}=\frac{d}{N_{%
\mathop{\rm eff}%
}}\sim 0.46\text{ nm},  \label{eq7}
\end{equation}
which is not far from the $c$-axis spacing $c=0.3524$ nm \cite{Nature}.
Thus, we may conclude that the description of superconducting fluctuations
above $T_{c}$ in the framework of a two-dimensional model is consistent with
our experiment.

In summary, we have measured the resistance and critical fields of MgB$_{2}$
films prepared on MgO substrate. From the linear temperature dependence of
the critical fields we established anisotropic three-dimensional
superconductivity of our films. From the resistivity measurements we showed
that the temperature dependence of fluctuation conductivity above $T_{c}$
agrees with the Aslamazov-Larkin theory for $2D$ superconductor. We consider
this finding as an experimental evidence for two-dimensional nucleation of
superconductivity in MgB$_{2}$.

\section{Acknowledgments}

The authors thank M. Klemm for helpful comments concerning the sample
structure characterization. This work was partially supported by INTAS Grant
No. 99-00585\ (A.S.S.).

\begin{center}

{\bf Figure captions}
\end{center}

Fig. 1. Resistive transitions, $R(T),$ for a 30 nm thick MgB$_{2}$ film at
different values of the DC magnetic field $B_{\perp }$: 0 - 0T, 1 - 0.3T, 2
- 0.6T, 3 - 0.9T, 4 - 1.8T, 5 - 2.7T, 6 - 3.6T, 7 - 4.5T, 8 - 5.4T, 9 - 6.9T.

Fig. 2. The temperature dependences of the parallel critical magnetic field $%
B_{c2\parallel }(T)$ (solid circles) and perpendicular critical magnetic
field $B_{c2\perp }(T)$ (solid squares) obtained from the midpoints of $R(T)$
data of Fig. 1.

Fig. 3. The temperature dependence of the inverse fluctuation conductivity, $%
\left[ \sigma ^{\prime }(T)\right] ^{-1},$ normalized by the normal state
value, $\sigma _{n},$ for the resistive transition in zero field (data of
curve ''0'' from Fig. 1). The straight line is the fit of experimental data
by Eq. (\ref{eq2}) for $2D$-fluctuations, and the curve is the fit for the $%
3D$-case.

\end{document}